\documentclass[twocolumn,showpacs,aps,amsmath,amssymb,amsfonts,floatfix]{revtex4}

\usepackage{graphicx}
\usepackage{dcolumn}
\usepackage{bm}
\usepackage{colordvi}
\usepackage{float}

\begin{document}

\preprint{APS/123-QED}

\title{Shear banding and flow-concentration coupling in colloidal glasses}

\author{R.~Besseling$^1$, L.~Isa$^1$, P.~Ballesta$^2$, G.~Petekidis$^2$, M.~E.~Cates$^1$ and W.~C.~K.~Poon$^1$}

\affiliation{$^1$SUPA, School of Physics \& Astronomy, The University of Edinburgh, \\
Kings Buildings, Mayfield Road, Edinburgh EH9 3JZ, United Kingdom.\\
$^2$ IESL-FORTH and Department of Materials Science and Technology, University of Crete, Heraklion 71110, Crete,
Greece.}


\begin{abstract}

We report experiments on hard sphere colloidal glasses that show a
type of shear banding hitherto unobserved in soft glasses. We
present a scenario that relates this to an instability due to
shear-concentration coupling, a mechanism previously thought
unimportant in these materials. Below a characteristic shear rate
$\dot\gamma_c$ we observe increasingly non-linear and localized
velocity profiles. We attribute this to very slight concentration
gradients in the unstable flow regime. A simple model accounts for
both the observed increase of $\dot\gamma_c$ with concentration,
and the fluctuations in the flow.

\end{abstract}



\maketitle

\noindent Shear banding is widespread in the flow of
disordered materials, including complex fluids \cite{fluidbanding}; pastes, gels and emulsions
\cite{OvarlezRheoAct09}; granular matter \cite{FenisteinNat03},
soils and rocks \cite{RudnickiJMPS75_banding} and metallic glasses
\cite{MGdilataion_model_exp}. Understanding this phenomenon is thus
crucial in various fields of science and engineering.

Constitutive models of shear banding exist, e.g. in wormlike
micellar fluids \cite{fluidbanding}, where a flow curve of stress
versus strain rate $\sigma(\dot\gamma)$ with slope $\eta_d<0$
causes instability and separation into two bands with distinct
flow rates $(\dot\gamma_1, \dot\gamma_2)>0$ \cite{fluidbanding};
these can also have distinct concentrations \cite{Wagner10}. Many
systems with a yield stress ($\sigma_y>0$) also show coexistence
of distinct bands
\cite{RogersVlassopoulosPRL08_agingbanding,Moller_lambdamodel,CoussotPRL02_shearbanding,BertolaJRheo03_emulsionslipyield},
with one band now being solid ($\dot\gamma_1 =0$). In some cases
this stems from a similar mechanical instability ($\eta_d<0$ for
$\dot\gamma<\dot \gamma_2$)
\cite{PicardFielding,Moller_lambdamodel,BertolaJRheo03_emulsionslipyield,CoussotBonnPRL02_avalanche}
due to positive feedback between flow and structural breakup.
However the banding seen in other experiments
\cite{expbanding,katgert_epl10,DivouxMannevillePRL10,Goyon} and
simulations \cite{VarnikPRL03_LJbanding,ShiFalkPRL07_STZbanding},
cannot be explained in those terms, particularly for purely
repulsive interactions \cite{MCT}. In some of these systems,
banding may be attributed to cooperativity between local plastic
events, characterized by a cooperativity length $\xi$. However,
theory \cite{BocquetPRL09_KEPmodel} and simulations
\cite{microlength} show that $\xi$ grows when the flow slows down,
which contrasts with the rate-independence seen in experiments
\cite{Goyon}.

In this Letter we show by experiment and theory that concentrated
hard-sphere (HS) colloids, one of the simplest yield-stress fluids
and a model for soft glasses generally, can exhibit a type of
shear banding that does not fit into any of the above categories.
Instead, we propose a scenario where banding is caused by
shear-concentration coupling (SCC). Though well known as a generic
mechanism for flow instability \cite{SchmittPRE95_banding}, this
has not previously been explored as a shear banding mechanism in
glasses. This is perhaps because the concentration changes
involved can be extremely small, as we show below; hence they are
not directly detectable in experiments. Crucially, the effects on
flow of very small concentration gradients are vastly amplified by
the presence of a yield stress.

It is well known that for {\it nonuniform} stress, particle
migration \cite{NottBradyJFM94} is driven by gradients in
$\dot\gamma$ and the nonequilibrium particle pressure,
$\Pi(\dot\gamma)$
\cite{YurkoMorrisJRheo08_particlepressure,DeBoeufPRL09_shear_Posm}.
The resulting concentration inhomogeneity causes departures from
flow profiles for the homogeneous system. But this can not explain
such departures under {\it uniform} stress, which requires {\it
intrinsic} instability \cite{SchmittPRE95_banding} (see also
\cite{NottBradyJFM94}). Recent theories \cite{furukawa} show that
Newtonian and linearly viscoelastic materials can also exhibit
such instability via SCC. However, this arises at 45$^\circ$ to
the flow direction, and is unrelated to the results in
\cite{SchmittPRE95_banding} on SCC-induced instability in {\it
nonlinear} fluids. In applying the latter to glasses, our two key
ingredients are (non-linear) dilatancy -the tendency of jammed
systems to expand under flow, ($\partial\Pi/\partial\dot\gamma>0$)
and flow nonlinearity. As both are ubiquitous in glassy materials,
our results are likely to have wide relevance.

We used sterically stabilised polymethylmethacrylate particles
(radii $a=138$~nm and $150~$nm, polydispersity $\sim 15\%$)
suspended in a decalin-tetralin mixture (viscosity
$\eta_s=2.3$~mPas) and seeded with $\sim 0.5\%$ fluorescent
colloids ($a=652$~nm) of the same kind. Different volume fractions
$\phi$ were prepared by diluting samples centrifuged to a sediment
with $\phi=\phi_m \simeq 0.67$; we report data in terms of the
reduced concentration $\Phi=\phi/\phi_{m}$. The glass transition
was found to lie at $\Phi_g \simeq 0.86$
\cite{MegenPRE98_tracersinglass}.

Rheology was measured in an AR2000 rheometer in cone/transparent
plate geometry (cone angle $1^{\circ}$, radius $20$~mm), coupled
to a confocal microscope \cite{BallestaPRL08,BesselingACIS09} to
measure the velocity $v(z)$ across the gap, $0 \leq z \leq z_{g}$,
at various $z_g$. For $\dot\gamma \gtrsim 0.1~$s$^{-1}$ we
measured $v(z)$ at each $z$ for a time $\geq 3/\dot\gamma$; for
$\dot\gamma \lesssim 0.1~$s$^{-1}$, $v(z)$ was measured from rapid
3D scans \cite{BesselingACIS09} over a time $\sim 2/\dot\gamma$.
The reported $v(z)$ are constant over the measurement time. To
prevent slip \cite{BallestaPRL08}, walls were coated with a
disordered monolayer of the tracers; results for a cone and/or
plate roughened to $\sim 10~\mu$m were very similar. We took data
at fixed $\dot\gamma$, stepping down $\dot\gamma$ after $300~$s
preshear at $\dot\gamma \simeq 10-30~$s$^{-1}$ or stepping up from
$\dot\gamma=0$ (waiting for $\geq 2/\dot\gamma$ before acquiring
$v(z)$ in either case). The results showed no systematic
differences between these protocols; stress-controlled
measurements also showed no significant changes.

\begin{figure}[t]
\includegraphics[width=0.45\textwidth,clip]{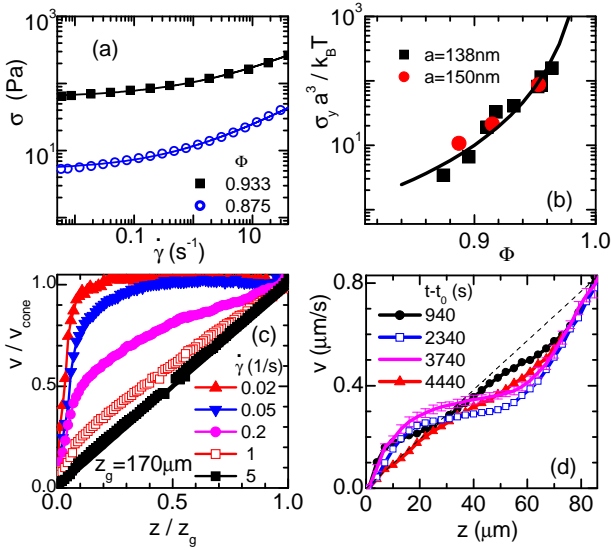}
\vspace{-0.0cm} \caption{(a) Flow curves $\sigma(\dot\gamma)$ with
HB fits (lines) for $\Phi=0.875$ and $\Phi=0.933$. (b) Yield
stress versus $\Phi$. Line: $\sigma_y=\sigma_0(1-\Phi)^{-3}$ with
$\sigma_0= 0.01k_BT/a^3$. (c) Velocity profiles $v(z)$ for
$\Phi=0.933$ at various $\dot\gamma$. (d) Evolution of $v(z)$
after startup shear of $\dot\gamma =0.01~$s$^{-1}$ at $t_0$
($\Phi=0.94$); data for $t-t_0=3740~$s show error bars. Data in
(a),(c) and (d) are for $a=138~$nm.} \vspace{-0.0cm} \label{fig:1}
\end{figure}

For $\Phi \gtrsim \Phi_g$, the bulk rheology is as previously
reported, Fig.~\ref{fig:1}(a), with flow curves of
Herschel-Bulkley (HB) form: $\sigma-\sigma_y \propto
\dot\gamma^n$, with $n\simeq 0.4-0.5$. The strong increase of
$\sigma_y$ with $\Phi$, due to the vanishing of free volume as
$\Phi \rightarrow 1$ \cite{fn_sigydiv}, is consistent with
$\sigma_y(\Phi)\simeq \sigma_0(1-\Phi)^{-p}$ with $p\simeq 3$ and
$\sigma_0 \simeq 0.01k_B T/a^3$, Fig.~\ref{fig:1}(b).

\begin{figure}[t]
\includegraphics[width=0.45\textwidth,clip]{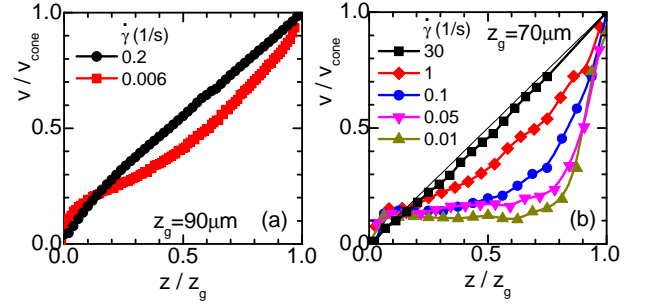}
\vspace{-0.0cm} \caption{Velocity profiles at different volume
fractions: (a) $\Phi=0.895$ ($a=138~$nm), (b) $\Phi=0.948$
($a=150~$nm).} \vspace{-0.0cm} \label{fig:2}
\end{figure}

Until now, HB and similar monotonic flow curves for glasses have
not been linked to non-transient shear banding. However, the
underlying velocity profiles, shown in Fig.~\ref{fig:1}(c) for
$z_g=170~\mu$m and $\Phi=0.933$, exhibit a marked change when we
decrease the imposed shear rate
$\dot\gamma=\int\dot\gamma(z)dz/z_g$ (here $\dot\gamma(z) \equiv
\partial_z v$). At large $\dot\gamma$, $v(z)$ is linear, but for
$\dot\gamma=0.2~$s$^{-1}$ $v(z)$ becomes highly nonlinear, with an
enhanced rate near the plate and a progressive reduction towards
the cone. For even smaller rates, $\dot \gamma \leq
0.05~$s$^{-1}$, the nonlinearity grows and $\dot\gamma(z)$
decreases {\em continuously} from a value $\gg \dot\gamma$ near
the plate to $\dot\gamma(z)\simeq 0$ for larger $z$. The width of
the fluidized band appears to saturate for low $\dot\gamma$ at
$\sim 80 a$ (not shown); we find no evidence for a minimum strain
rate in this band \cite{BesselingPRL2007}. The behavior for
$z_g=60~\mu$m and $90~\mu$m is essentially the same. Such
continuously varying flow profiles strongly contrast with the
distinct solid and fluid bands in thixotropic yield stress fluids
\cite{Moller_lambdamodel,BertolaJRheo03_emulsionslipyield}. HS
glasses, which show only very weak aging of quiescent properties
\cite{fn_sigydiv}, are thus distinct from such systems. Note from
Figs.~\ref{fig:1}(c) and~\ref{fig:2}(b) that $v(z)$ has no unique
`symmetry': the fluidized band may appear near either the plate or
the cone. This rules out sedimentation or specific wall rheology
\cite{Goyon} as explanations \cite{fn_stressnonuni}. The growth of
the fluidized band with $\dot\gamma$ also contrasts with the rate
dependence (or lack thereof) of the cooperativity length found in
\cite{BocquetPRL09_KEPmodel,Goyon}.

\begin{figure}
\includegraphics[width=0.4 \textwidth]{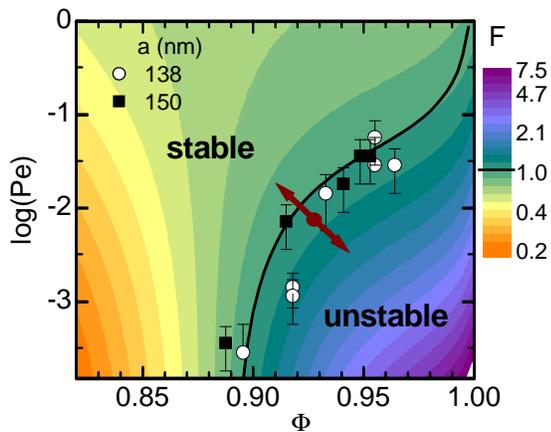}
\vspace{-0.0cm} \caption{Line: critical flow rate ${\rm
Pe_c}(\Phi)$ from Eq.~(\ref{eq:Pe_c_low}) with
$[p,r,m=n,A,B]=[3,4,0.4,25,0.005]$.
Colors mark the value of $F$. Symbols: experimental values for
${\rm Pe_c}$ (averaged over $z_g$). Arrow: possible evolution of
an unstable state.} \vspace{-0.0cm} \label{fig:3}
\end{figure}

Next, we discuss the concentration dependence of the observed
behavior, Fig.~\ref{fig:2}. For both $\Phi=0.895$ (just within the
glass, Fig.~\ref{fig:2}(a)) and for a much higher concentration
$\Phi=0.948$, Fig.~\ref{fig:2}(b)), we again observe a transition
to a nonlinear velocity profile as $\dot\gamma$ is lowered, but
the shear rate at which this occurs is respectively much smaller
and higher than for $\Phi = 0.933$, Fig.~\ref{fig:1}(c). On
decreasing $\dot\gamma$, we define the critical shear rate,
$\dot\gamma_c$, to be that point at which the maximum deviation
from linearity of the normalized velocity profile,
$|v(z)-\dot\gamma z |/v_{\rm cone}$, first exceeds 0.1. Results
for the critical P\'{e}clet number, ${\rm Pe_c}=\dot\gamma_c
\tau_B$ (where $\tau_B=6\pi \eta_s a^3/k_B T$ is the Brownian
time) are shown as a function of $\Phi$ in Fig.~\ref{fig:3}.

\begin{figure}[t]
\includegraphics[width=0.45\textwidth,clip]{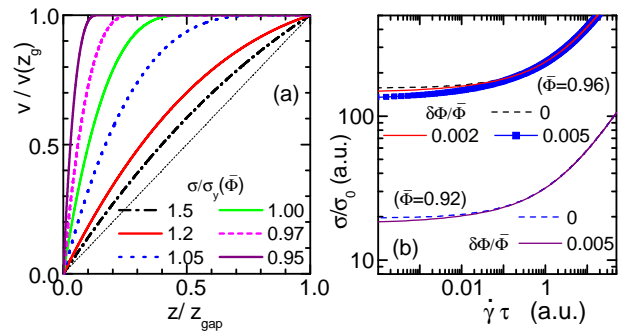}
\vspace{-0.0cm} \caption{Velocity profiles $v(z)/v(z_{g})$,
calculated from $\dot\gamma \propto
[(\sigma/\sigma_y(\Phi(z)))-1]^{2}$ and $\sigma_y \propto
(1-\Phi)^{-3}$, with
$\Phi(z)-\bar{\Phi}=|\delta\Phi|[(z/z_{g})-(1/2)]$ with
$\bar{\Phi}=0.96$, $|\delta\Phi|/\bar{\Phi}=0.002$, for different
$\sigma/\sigma_y(\bar{\Phi})$. (b) Effective flow curves $\sigma$
versus $\dot\gamma=[v(z_{g},\sigma,\Phi(z))]/z_{g}$ for various
$\bar{\Phi}$ and $|\delta \Phi|$. \vspace{-0.0cm} \label{fig:4}}
\end{figure}

To begin to interpret our observations, we first show that the
observed velocity profiles can be reconciled with HB behavior,
simply by postulating a small concentration variation $\delta
\Phi(z)$ across the gap. Writing the HB form as $\dot\gamma \tau=
[(\sigma/\sigma_y(\Phi))-1]^{1/n}$ with $\sigma_y(\Phi) =
\sigma_0(1-\Phi)^{-3}$ as before, we can then calculate
$v(z,\sigma)=\int_0^z \dot\gamma(\bar{\Phi}+\delta
\Phi(z'),\sigma)dz'$ for a given mean concentration $\bar{\Phi}$
and a choice of $\delta\Phi(z)$. In Fig.~\ref{fig:4}(a) we do this
for a uniform gradient $\partial \delta \Phi(z)/\partial z =
|\delta\Phi|/z_g$ with $|\delta\Phi|/\bar{\Phi}=0.002$, at various
values of the reduced stress $\sigma/\sigma_y(\bar\Phi)$. When
$\sigma$ approaches $\sigma_y(\bar\Phi)$, $v(z)$ changes from
weakly to strongly nonlinear, reflecting the progressive
localization of shear within regions of the sample with the lowest
yield stress $\sigma_y(\Phi(z))$, i.e. with the lowest $\Phi(z)$.
These results strikingly resemble the experimental data in
Fig.~\ref{fig:1}(b), although $\delta\Phi$ is too small to be
directly measured \cite{fn_deltaphi}; different symmetries of
$v(z)$ in other experiments can also be explained by corresponding
changes in $\delta\Phi(z)$. Note that the mean shear rate
$\dot\gamma=v(z_{g},\sigma,\Phi(z))/z_{g}$ differs from
$\dot\gamma(\sigma,\bar{\Phi})$, but the {\it effective} flow
curves $\sigma(\dot\gamma,\Phi(z))$ deviate only slightly from the
uniform $\sigma(\dot\gamma,\bar{\Phi})$, see Fig.~\ref{fig:4}(b).

While concentration gradients can thus account for the results,
their origin and the enhanced shear localization with increasing
$\bar\Phi$ remain to be explained. We now show that both are
explicable via the SCC instability scenario of
\cite{SchmittPRE95_banding}. Fluctuations in concentration
($\delta\Phi$) and shear rate ($\delta\dot\gamma$) evolve via the
diffusion and Navier-Stokes equations, in which shear-induced
migration and the $\Phi$ dependence of the shear stress must be
included \cite{SchmittPRE95_banding}. For small fluctuations along
$z$ we have (to linear order in $\delta\Phi$, $\delta\dot\gamma$):
\begin{eqnarray}
\partial_t\delta\Phi=-\vec{\nabla}\cdot\vec{J}\simeq&
M\left(
\Pi_{\Phi}~\partial^2_z\delta\Phi+
\Pi_{\dot\gamma}~\partial^2_z\delta \dot\gamma \right),
\label{eq:lin1}\\
\partial_t\delta \dot\gamma=\rho^{-1}\partial_z^2
\sigma  \simeq& \rho^{-1}\left(
\sigma_{\Phi}~\partial^2_z\delta\Phi +
\sigma_{\dot\gamma}~\partial^2_z\delta\dot\gamma\right).\label{eq:lin2}
\end{eqnarray}
Here we have introduced the shorthand $\Pi_{\dot\gamma} \equiv
\partial\Pi/\partial\dot\gamma|_{\Phi}$, likewise
$\Pi_\Phi,\sigma_{\dot\gamma}$ and $\sigma_\Phi$; $M$ is a
collective mobility and $\rho$ the density. The migration current
$\vec{J}$ arises from particle pressure gradients $\partial_z
\Pi(\Phi,\dot\gamma)$ due to variations in both $\Phi$ and
$\dot\gamma$ \cite{YurkoMorrisJRheo08_particlepressure,fn_mu}. The
terms involving $\Pi_{\dot\gamma}$ and $\sigma_\Phi$ in
Eqs.~(\ref{eq:lin1},\ref{eq:lin2}) cause respectively particle
migration towards regions of low shear rate, and accelerated shear
in regions of low concentration; together, these amplify
fluctuations. That is, a fluctuation towards higher $\Phi$ in some
region creates a lower shear rate there. This promotes inward
particle migration, giving a positive feedback effect. This
tendency is counteracted by the remaining terms which describe
stable diffusive spreading of both particles and momentum (or
equivalently shear rate). Rewriting
Eqs.~(\ref{eq:lin1},\ref{eq:lin2}) as $\partial_t\Psi_i =
L_{ij}\partial_z^2\Psi_j$ with $\Psi_i =
(\delta\Phi,\delta\dot\gamma)$, we see that instability sets in
when $\det L_{ij} =
M(\Pi_\Phi\sigma_{\dot\gamma}-\Pi_{\dot\gamma}\sigma_\Phi)/\rho$
becomes negative, or equivalently, as first derived in
\cite{SchmittPRE95_banding,fn_mu}, when
\begin{equation}
F\equiv \frac{\Pi_{\dot\gamma}\sigma_\Phi}{\Pi_\Phi
\sigma_{\dot\gamma}}>1. \label{eq:modSchmittcrit}
\end{equation}
To evaluate $F$, we first write the HB form for $\sigma$ in terms
of the P{\'e}clet number ${\rm Pe}=\dot\gamma \tau_B$:
\begin{equation}
\sigma=\frac{\sigma_0}{(1-\Phi)^{p}}[1+s(\Phi){\rm Pe}^n],
~~s(\Phi)=A(1-\Phi)^n. \label{eq:sigmaform}
\end{equation}
The first term is $\sigma_y$ and $n\simeq 0.4-0.5$, as before.
Typical values for $A$ from our experiments are $A=10-20$. The
particle pressure $\Pi$ for HS colloids has a similar form:   
\begin{equation}
\Pi=\frac{\Pi_0\Phi}{(1-\Phi)}[1+g(\Phi){\rm
Pe}^m],~~g(\Phi)\equiv B(1-\Phi)^{1-r}. \label{eq:piform}
\end{equation}
Here $\Pi_0=2.175\phi_m k_BT/\pi a^3$; the first term then
approximates the osmotic pressure at rest
\cite{BradyJrheo95_normalstress}, whereas the second term is the
contribution due to shear. For the latter, a ${\rm Pe}^2$
dependence was found in
\cite{BradyJrheo95_normalstress,FossBradyJRhe02000}, but this is
restricted to a linearly viscous regime ($\sigma \propto {\rm
Pe}$) at very small ${\rm Pe}$, in which
Eq.~\ref{eq:modSchmittcrit} yields $F \propto {\rm Pe}^2$
resulting in stable flow. However, this regime is hard to access,
and indeed completely absent whenever $\sigma_y$ is nonzero
($\Phi>\Phi_g$), where it is replaced by a non-Newtonian regime in
which both the stress and the particle pressure increase
sublinearly with ${\rm Pe}$. From simulations for $\Phi\lesssim
\Phi_g$
\cite{FossBradyJRhe02000,YurkoMorrisJRheo08_particlepressure}, we
extract $m=0.4-0.5$, (and $B \simeq 0.003$, $r \simeq 3$), so that
$m \simeq n$, as is also observed in simulations of 2D foams
\cite{LangloisWeariePRE08_foamsim}. For $\Phi \gtrsim \Phi_g$, and
for glassy flow in general, we expect these two exponents to
remain similar ({\em e.g.}, within mode coupling theory, $m=n$
seems probable \cite{MCT}).

Using Eqs.~(\ref{eq:sigmaform},\ref{eq:piform}), we obtain a
limiting value for $F$ at large ${\rm Pe}$ as
$F_{\infty}=m(p-n)/nr$ \cite{fn_approx}. The flow is thus stable
in this regime provided $r
> p-n$; to explain our experiments we require $r \gtrsim 2.5$.
More interesting is the result for small ${\rm Pe}$, where we
obtain
\begin{equation}
F \to {\rm Pe}^{m-n}\, \frac{mpg(\Phi)\Phi }{ns(\Phi)}\simeq
\frac{mpB\Phi}{nA(1-\Phi)^{r+n-1}} = F_0. \label{eq:F0}
\end{equation}
Here $F_0(\Phi)$ is a quasi-plateau value maintained while ${\rm
Pe}^{m-n}\simeq 1$ (and a true limiting value if $m=n$). It
follows that homogeneous flow at low ${\rm Pe}$ is {\em unstable}
for concentrations $\Phi > \Phi_c$, where $F_0(\Phi_c) = 1$. We
argue that it is this SCC-induced instability that creates the
small variations $\delta\Phi(z)$ that were assumed in
Fig.~\ref{fig:4}(a), and which account for the experimentally
observed localized flow at small $\dot\gamma$. Because, as in
\cite{SchmittPRE95_banding}, our analysis is limited to linear
stability, we do not have a clear idea of the mechanism limiting
the growth of $\delta \Phi$. However, the extreme nonlinear
dependence of both $\sigma$ and $\Pi$ on $\Phi$ as $\Phi\to 1$
makes it plausible that $\delta\Phi$ remains small.

The critical flow rate ${\rm Pe_c}$, below which the instability
sets in, follows from the condition $F({\rm Pe_c},\Phi)=1$. For
the case $m=n$ this simplifies to \cite{fn_approx}:
\begin{eqnarray}
{\rm Pe_c}(\Phi) \simeq \left[\frac{p}{r
s(\Phi)(1-F_{\infty})}\left(1-\frac{1}{F_0(\Phi)}
\right)\right]^{1/n}. \label{eq:Pe_c_low}
\end{eqnarray}
Figure~\ref{fig:3} shows ${\rm Pe}_c(\Phi)$ for specific parameter
values. Entering the unstable regime $\Phi>\Phi_c$, ${\rm Pe}_c$
increases and grows $\sim s(\Phi)^{-1/n}$ for $\Phi \to 1$. Our
model is able to give a quantitative account of the data; the fit
\cite{fn_fitting} gives $\Phi_c = 0.89$, close to, but above,
$\Phi_g$. (Note that Eq.~(\ref{eq:F0}) does not rule out
$\Phi_c<\Phi_g$ in principle.)

In contrast to most shear banding scenarios \cite{fluidbanding},
the present system seems unable to achieve global stability by
separating into distinct bands. This is illustrated by the arrows
in Fig.~\ref{fig:3}: an initially unstable state can form a
locally depleted region, $\Phi(z)<\bar{\Phi}$, that is stabilized
when ${\rm Pe}(\Phi(z))> {\rm Pe}_c$, but the remaining
concentrated region is even more unstable than before. This
suggests that the banded flow should have residual temporal
fluctuations. We have indeed observed this (Fig.~\ref{fig:1}(d));
after startup shear of $\dot\gamma=0.01~$s$^{-1}$, a weakly
nonlinear profile develops a central region where $\dot\gamma(z)$
is strongly reduced, which then speeds up by expanding the lower
band, and subsequently reverts to a larger $\dot\gamma(z)$. We
have also observed (in a planar shear cell
\cite{BesselingPRL2007}) that a fluidized band can swap from $z
\simeq 0$ to $z \simeq z_g$ over sufficient time. Leaving a
detailed study of these effects for future work, we note that the
shear banding we observe is permanent (if unsteady) not transient;
moreover, our stress is time-independent, in contrast to the
results in \cite{DivouxMannevillePRL10}.

Our shear-banding mechanism arises from the concentration
dependent nonlinear rheology of glasses along with the nonlinear
process of `Brownian dilation' whereby flow increases the particle
pressure $\Pi(\Phi,{\rm Pe})$. For larger ${\rm Pe}$ than those
studied here, hydrodynamic effects lead to a much stronger, linear
increase of $\Pi$ with flow rate \cite{DeBoeufPRL09_shear_Posm},
ultimately leading to shear thickening in these suspensions.
Hence, the localization we observe might be interpretable as a
precursor to shear thickening.


In conclusion, we have shown that HS colloidal glasses exhibit a
new type of shear banding, well described by a model in which SCC
leads to unstable flow near yielding. In this scenario, very small
concentration variations can sustain large variations in flow
rate. Our results may also be relevant for flow in other glassy
materials, such as foams where `dilatancy' has recently been
observed \cite{MarzeColSurfA05}. It may also shed light on shear
bands and dilatancy in metallic glasses
\cite{MGdilataion_model_exp}. We mention in particular simulations
of a model glass \cite{VarnikPRL03_LJbanding}, which showed both
fluctuating shear bands and small but finite concentration
gradients.

We thank A.B. Schofield for colloids, and J. Brader, P. Coussot,
D. Marenduzzo, A. Morozov, H. Tanaka, M. van Hecke and T.
Voigtmann for discussions. Work funded in part by EPSRC EP/D067650
and EP/E030173. L.I. was funded by EU MRTN-CT-2003-504712. G.P.
and P.B. acknowledge EU funding from ToK `Cosines'
(MTCD-CT-2005-029944) and NMP Small `Nanodirect'
(CP-FP7-213948-2). M.E.C. is funded by the Royal Society.


\begin{thebibliography}{}
\bibitem{fn_stressnonuni}In our geometry $\sigma(z)$ is uniform within
$0.03~\%$.
\bibitem{fn_deltaphi}The typical $|\delta\Phi|/\bar{\Phi}$ needed to mimic $v(z)$ in
experiments corresponds to $\delta\phi\lesssim 0.3\%$, below our
confocal imaging resolution ($\gtrsim 0.5\%$). MRI experiments as
in \cite{OvarlezPRE08_emsulsionshear} may be able to detect this.
The associated change in the measured $\sigma_y$ is $\delta
\sigma_y/\sigma_y(\bar{\Phi})
\simeq(p|\delta\Phi|/2)/(1-\bar{\Phi}) \lesssim 10\%$. Due to the
related experimental uncertainty in $\sigma_y(\bar{\Phi})$, it is
dificult to extract $\delta\Phi(z)$ from the measured $v(z)$ near
$\sigma_y$.
\bibitem{fn_mu}In \cite{SchmittPRE95_banding}, $J$ was described via a chemical potential
$\mu(\Phi,\dot\gamma)$.
Here we use $J = -M\left[
\partial_{\Phi}\Pi~\nabla \Phi+\partial_{\dot\gamma}\Pi~\nabla
\dot\gamma \right]$ as in
\cite{DeBoeufPRL09_shear_Posm,YurkoMorrisJRheo08_particlepressure}.
\bibitem{fn_approx}We have used the fact that $1-\Phi \ll r\Phi$.
\bibitem{fn_fitting}In the fit we fix $p=3$, $m=n$, and vary $A, B, n$ and $r$.
\end{thebibliography}
\end{document}